\documentclass[lettersize,journal]{IEEEtran}
\usepackage{amsmath,amsfonts}
\usepackage{algorithmic}
\usepackage{array}
\usepackage[caption=false,font=normalsize,labelfont=sf,textfont=sf]{subfig}
\usepackage{textcomp}
\usepackage{stfloats}
\usepackage{url}
\usepackage{verbatim}
\usepackage[table,xcdraw]{xcolor}
\usepackage{graphicx}
\usepackage{hyperref}
\usepackage{caption}
\def\BibTeX{{\rm B\kern-.05em{\sc i\kern-.025em b}\kern-.08em
    T\kern-.1667em\lower.7ex\hbox{E}\kern-.125emX}}
\usepackage{balance}
\usepackage{amsmath,amssymb,amsfonts}

\usepackage{float}
\usepackage{xcolor}
\usepackage{algorithmic}
\usepackage{mathtools}

\usepackage{textcomp}
\usepackage{xcolor}
\usepackage{tabularx,booktabs}
\usepackage{varioref}
\usepackage{Keywords}
\usepackage{tikz-cd}
\usepackage{soul}
\newcolumntype{C}{>{\centering\arraybackslash}X} 

\begin{document}

\title{A Stochastic Differential Equation Framework for Modeling Queue Length Dynamics Inspired by Self-Similarity}

\author{Shakib~Mustavee\textsuperscript{1},
        ~Shaurya~Agarwal\textsuperscript{1$^{\dagger}$} (Senior Member IEEE), 
        ~Arvind~Singh\textsuperscript{1}
\thanks{$^{1}$ The authors are with the Department of Civil Engineering, University of Central Florida, Orlando, USA }
\thanks{$^{\dagger}$ Corresponding Author}}

\maketitle




\begin{abstract}
This article develops a stochastic differential equation (SDE) for modeling the temporal evolution of queue length dynamics at signalized intersections. Inspired by the observed quasiperiodic and self-similar characteristics of queue length dynamics, the proposed model incorporates three properties into the SDE: (i) mean reversion with periodic mean, (ii) multiplicative noise, and (iii) fractional Brownian motion. It replicates key statistical features observed in real data, including the probability distribution function (PDF) and PSD of queue lengths. This is, to our knowledge, the first equation-based model for queue dynamics. The proposed approach offers a transparent, data-consistent framework that may help inform and enhance the design of black-box learning algorithms with underlying traffic physics.

\end{abstract}
 



\section{Introduction}\label{sec:intro}
Accurately modeling traffic flow along signalized corridors is inherently challenging due to the complex dynamics that arise from interactions among interconnected intersections. The system’s high dimensionality, coupled with its stochastic variability, presents significant challenges for both analytical formulations and computational modeling approaches \cite{li2025traffic}. Macroscopic variables that describe traffic states on signalized corridors include average delay, queue length, average speed, flow rate, and density \cite{maripini2023traffic}. Accurate modeling of these variables is essential for the efficient management and control of signalized arterials. Among these variables, queue length is often regarded as a critical measure of effectiveness (MOE) in adaptive signal timing control algorithms \cite{zheng2019diagnosing}. Consequently, modeling queue length dynamics plays a crucial role in managing traffic flow along adaptive corridors. Queue length dynamics typically describe how the maximum vehicle accumulation during the red phase at a given approach evolves across signal cycles. Understanding this temporal evolution improves prediction capabilities and enhances decision-making. As a result, the topic has received considerable attention from the research community. Queue length models can broadly be categorized into two groups: (i) theoretical and (ii) data-driven approaches \cite{MEDINASALGADO2022100739}.


Theoretically, queue length is estimated from the conservation law. Traffic flow is analogous to the continuous flow of compressible fluids, governed by conservation laws. The LWR (Lighthill-Whitham-Richards) model describes the spatiotemporal evolution of traffic flow, assuming a linear relationship between traffic density and velocity in the conservation law. However, disruptions in the flow, such as lane closures, accidents, traffic bottlenecks, or signalized intersections, introduce discontinuities that result in shockwaves. Analyzing these shockwaves helps explain the queue formation process \cite{liu2009real}. Nevertheless, shockwave theory assumes an instantaneous transformation of flow, which can lead to errors in queue length estimation. The cell transmission model overcomes this limitation by discretizing the LWR model in both space and time, providing a more numerically efficient approach to state estimation \cite{zhao2015ctm}. Furthermore, state-space models, combined with filtering techniques, can be used for traffic state estimation, including queue lengths near signalized intersections \cite{yin2018kalman, abewickrema2023multivariate}. However, due to the highly stochastic and nondeterministic nature of traffic flow, the success of shockwave theory-based prediction models heavily depends on the accuracy and granularity of input data, making the approach practically challenging \cite{gu2024cycle}. To address the limitations of theoretical or first-principle-based models, data-driven have gained popularity for modeling and predicting the temporal evolution of queue lengths at signalized arterials. Among these data-driven models, non-parametric approaches have been particularly successful in queue length prediction, with deep learning models showing remarkable performance. These models rely on historical patterns and do not depend on strict mathematical assumptions. Notable deep learning models for queue length prediction include LSTM \cite{rahman2021real}, spatial-temporal graph neural networks (STGNNs) \cite{shirakami2023qtnet}, temporal attention mechanisms \cite{sengupta2021tqam}, and geometric deep learning techniques \cite{wright2019neural, ehlers2019traffic}. Despite their success in prediction, these models often lack interpretability and typically involve a large number of parameters. In contrast, parametric models provide a fixed mathematical structure that is easier to interpret. These models are often used where transparency and simplicity are prioritized \cite{comert2021grey}.

\begin{figure}[htbp]
    \centering

    \subfloat[%
        \begin{footnotesize}
        Queue length time series (2-min resolution)
        \end{footnotesize}
        \label{fig:queue_sub}
    ]{%
        \includegraphics[width=0.9\linewidth]{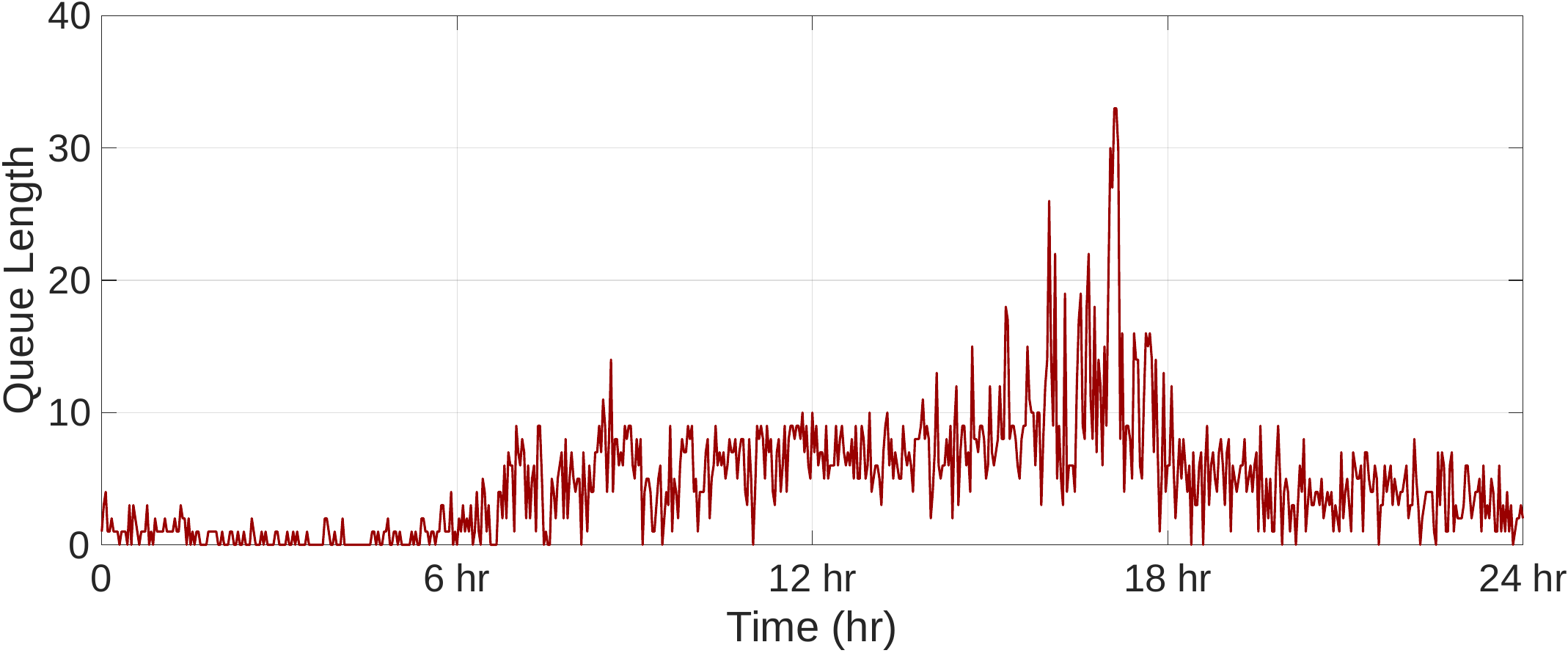}
    }\\[1ex]

    \subfloat[%
        \begin{footnotesize}
        Traffic flow time series (15-min resolution)
        \end{footnotesize}
        \label{fig:flow_sub}
    ]{%
        \includegraphics[width=0.9\linewidth]{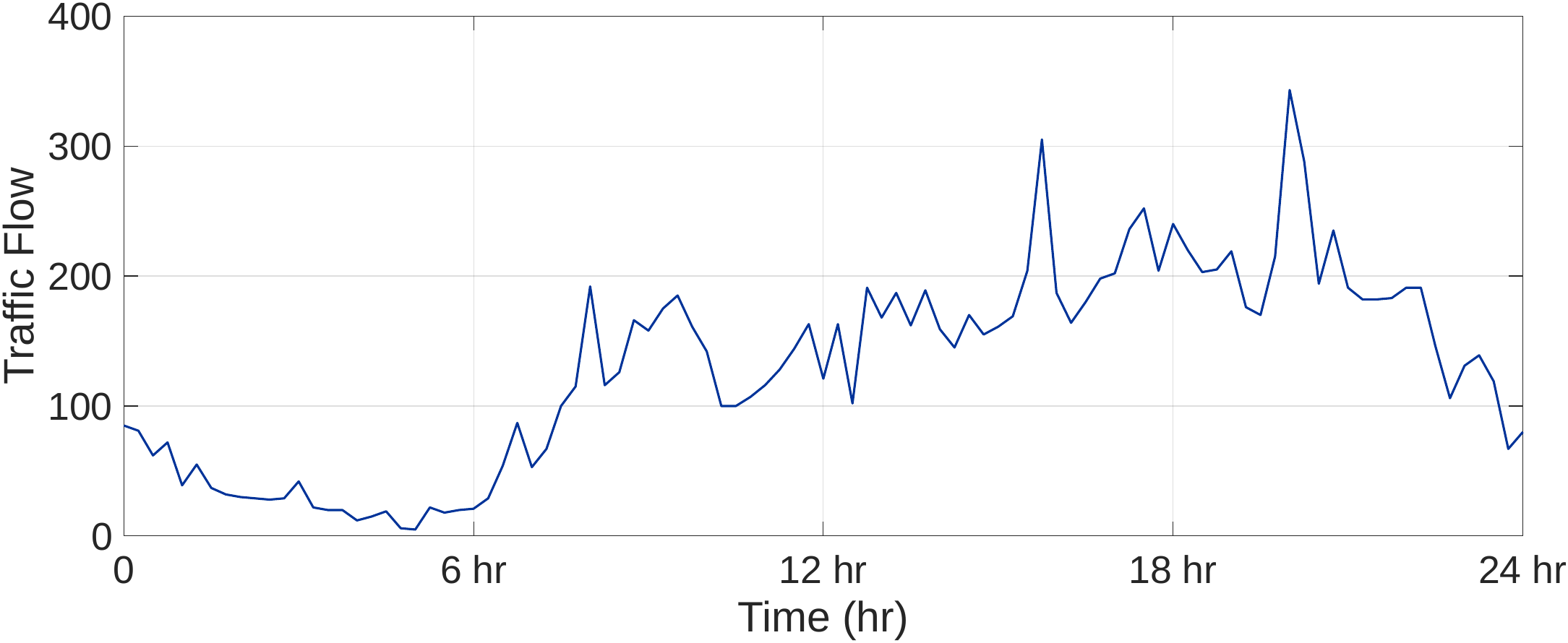}
    }

    \caption{Visualizing fluctuations in queue length and traffic flow on University Blvd. 
    Subfigure~1\textbf{(a)} shows the queue length with high fluctuation, while 
    subfigure~1\textbf{(b)} displays smoother traffic flow. The contrast suggests that stochastic 
    differential equations (SDEs) may be suitable for modeling queue dynamics.}
    \label{fig:flow_queue}
\end{figure}

Despite not being derived from first principles, parametric models have proven effective in capturing underlying data patterns and the interconnections between various traffic variables. These models are particularly useful in modeling the stochasticity and fluctuations observed in traffic data, especially in queue length time series. For example, queue length data is typically more granular, with cycle lengths of \(2\) to \(3\) minutes, whereas traffic flow data is usually aggregated over larger time windows, typically ranging from \(5\) minutes to \(1\) hour. This difference in aggregation windows is vital because larger windows tend to smooth out fluctuations, reducing the variability in the data and making its temporal evolution more stable \cite{vlahogianni2011temporal}. This difference in resolution is clearly illustrated in Figure~\ref{fig:flow_queue} (refer to Section~\ref{sec:empirical} for details), which shows queue length and traffic flow time series for the same signalized intersection on the same day. The queue length time series, with a \(2\) minute resolution, captures rapid fluctuations and variations in traffic flow that are essential for understanding the dynamics of queue formation and dissipation (Figure~\ref{fig:queue_sub}). On the other hand, the aggregated \(15\) minute resolution of the traffic flow time series smooths out these fluctuations, providing a less detailed view of the underlying traffic behavior (Figure~\ref{fig:flow_sub}). This difference underscores the importance of considering fluctuations and stochastic behavior in queue length dynamics. To the best of the authors' knowledge, there is no explicit mathematical equation for queue length dynamics in the literature. To address the research gap, this paper proposes a stochastic differential equation (SDE) to capture the inherent fluctuations and randomness in queue length dynamics. SDEs are flexible parametric models that can include additional terms to represent specific dynamic behaviors. The accuracy of the model relies on how well domain knowledge is integrated. Traffic time series often display long-range dependence \cite{feng2018better}, and self-similarity, which are typically characterized by a \(1/f\) -type power spectral structure \cite{feng2018better,yuan2017long,campari2002self}.


These properties were taken into account while formulating the proposed model to better reflect the real-world traffic behavior.



\noindent \textbf{Contributions:}
 The contributions of this article are diverse and multifaceted and are described as follows:

\begin{enumerate}
    
    \item Proposes a novel stochastic differential equation (SDE) model for queue length evolution, capturing long-term memory and fluctuations in real-world traffic systems.
    
    \item The proposed SDE is formulated by linking queue dynamics to microscopic traffic behavior, where vehicle interarrivals follow a point process, and headway dynamics is influenced by multiplicative noise.
    
    \item The probability density function (PDF) and power spectral density (PSD) of the simulated process using the proposed SDE is similar to those of real queue length time series.
    
\end{enumerate}

\noindent \textbf{Outline: }
The structure of the paper is as follows.  Section~\ref{sec:litrev} discusses the relevant literature on fractal analysis of traffic networks: Section~\ref{sec:empirical} explains the empirical studies while Section~\ref{sec:results_discussion} presents the proposed model, the simulation process, and the results. 

\section{Literature Review and Methods} \label{sec:litrev}
The literature review section will focus on two aspects first focusing on the existing literature regarding self-similarity or the \(1/f\)-type spectra observed in traffic flow. The section will discuss both freeway and signalized corridors and will try to provide a gradual development of this domain. This will help understand the literature gaps and future directions. Secondly, the paper will discuss the application of SDEs in modeling traffic flow. 

\subsection{Self-Similarity Analysis in Traffic Flow Dynamics }Here we include a brief history on the development of fractal theory in traffic flow theory will help the reader contextualize and grasp at the use of the theory in understanding traffic flow. The earliest observation of \(1/f\) spectra in traffic flow was made by Hinguich and Musha in expressway traffic flow in 1976. The authors reported \(1/f\) fluctuation in traffic density \cite{musha19761}.

\noindent \textbf{Theoretical Studies:} The theoretical analysis of self-similarity and \(1/f\) spectral characteristics in traffic flow has predominantly focused on cellular automata-based models. Takayasu and Takayasu \cite{takayasu19931} simulated a discrete, single-lane traffic model using cellular automata. They demonstrated a connection between \(1/f\) spectral behavior and phase transitions, showing that \(1/f\) fluctuations emerge in the temporal density during the jammed phase, under both stochastic and deterministic conditions. Choi \cite{choi1995traffic} examined traffic dynamics in the hydrodynamic limit, using the Navier–Stokes equation in conjunction with the continuity equation. The study classified traffic into laminar (homogeneous), chaotic, and density-wave phases, and revealed that current fluctuations in the density-wave regime exhibit \(1/f\) behavior in their PSD. In another significant contribution, Paczuski et al. \cite{paczuski1996self} analytically established a connection between spatial fractal properties and long-range temporal correlations. Their study highlighted that phantom jams, which result from intermittent stop-and-go patterns, possess a complex fractal structure. They derived the corresponding power spectral behavior, confirming its \(1/f\) nature. Nassab et al. \cite{nassab20051} explored the influence of on-ramp and off-ramp traffic on open-boundary roads using a cellular automata framework. Their simulations incorporated low-probability velocity fluctuations and produced evidence of \(1/f\) spectra in the resulting data. Further, Wu et al. \cite{wu2008long} analyzed the long-range dynamics of traffic using the Kerner–Klenov–Wolf (KKW) cellular automaton model. They applied Detrended Fluctuation Analysis (DFA) to study spatiotemporal complexity, average speed, and density. Their results showed long-range anticorrelations in free-flow and wide-moving jam phases, while synchronized flow exhibited long-range positive correlations. However, cellular automata models have limitations in fully explaining the fractal nature of traffic flow. For instance, Xue et al. \cite{xue2015long} examined synchronized traffic near a partially reduced two-lane bottleneck using the KKW model under Kerner’s three-phase theory. Through fundamental diagram analysis, cross-correlation, and rescaled range (R/S) methods, they found that synchronized flow displayed persistent long-range correlations, with Hurst exponents increasing as bottleneck length grew. In contrast, simpler models such as the NaSch model failed to replicate these long-range dependencies, emphasizing their inadequacy in modeling synchronized traffic regimes. \\

\noindent \textbf{Empirical Studies:} Empirical studies of self-similarity in traffic flow involves primarily time series analysis on real and simulated data. Apart from PSD analysis common mathematical tools used in these analysis are DFA (Detrended Fluctuation Analysis), MFDFA (Multifractal Detrended Fluctuation Analysis), MFDXA (Multifractal Detrended Cross-correlation Analysis), Rescaled Range (R/S) analysis, Autocorrelation Function (ACF), and H\"older \ exponent analysis on traffic flow speed, flow, and density time seires to characterize fractality behavior, reveal correlations between variables, and identify long-range dependence within traffic time series. 

P. Shang and collaborators conducted a series of studies applying time series-based techniques to analyze the fractal characteristics of freeway traffic flow. Using methods such as DFA, R/S analysis, and PSD analysis, they identified both monofractal and multifractal behaviors in traffic data. An increasing degree of fractality was reported with growing congestion levels, and the H\"older\ exponent was used to detect local fractal rates associated with congestion \cite{shang2006application}. In a separate study, the Hurst exponent was estimated to be \(0.84\), confirming long-range dependence in average traffic speed time series \cite{shang2007fractal}. Detrended fluctuation and cross-correlation analyses further revealed that traffic fluctuations exhibit power-law auto-correlations, while sign-separated fluctuations show anti-correlated behavior. Stronger cross-correlations were found between adjacent road sections than between adjacent lanes \cite{xu2010modeling}. Building on insights from monofractal analysis, multifractal behavior was explored in congested traffic, highlighting the complexity of speed fluctuations \cite{li2007multifractal}. Using MFDFA, Shang et al. demonstrated long-range dependence in high-resolution speed data \cite{shang2008detecting}, and later introduced a time-dependent Hurst exponent \(H(t)\) to quantify traffic variability over time. This analysis, based on real and synthetic data from Beijing’s Yuquanying highway, revealed pronounced temporal fluctuation and non-monotonic multifractal patterns \cite{yue2010time}. Additionally, multiscale multifractal analysis was employed to further understand traffic dynamics across different time scales \cite{wang2014multiscale}.

Several studies have analyzed urban and highway traffic flow through the lens of fractal and long-range dependence. Using detrended fluctuation analysis (DFA), Peng et al. characterized long-range correlations in urban traffic data from multiple Beijing intersections, revealing three-phase traffic characteristics akin to freeway traffic and showing that traffic restriction policies reduce both flow and irregular fluctuations \cite{peng2010long}. Krause et al. estimated the Hurst exponent in urban traffic and linked anti-persistent behavior to traffic jam durations, suggesting that traffic flow resembles fractional Brownian motion with time-varying fluctuations \cite{krause2017importance}. He et al. employed adaptive fractal analysis (AFA) and DFA to study traffic volumes on urban expressways, uncovering multifractal and long-range anti-persistent behavior on ramps, with more pronounced multifractality on weekdays \cite{he2016fractal}. Inspired by self-similarity in data network traffic, Meng et al. examined highway vehicle arrival patterns using real data from the Texas Department of Transportation, identifying clear evidence of self-similarity and deviations from Poisson assumptions \cite{meng2009self}. Similarly, Perati et al. investigated self-similar behavior at highway toll plazas and found that both mean queue length and busy period distribution increase with higher traffic flow and Hurst exponent \cite{perati2012self}.

Fractal analysis has been extensively applied to traffic flow from both theoretical and empirical perspectives. However, despite this rich body of research, its integration into practical transportation engineering applications—such as traffic control, design, or management—remains limited and underdeveloped \cite{laval2023self}.

\subsection{SDE in Traffic Time Series:} 


Stochastic differential equations (SDEs) offer a flexible framework for capturing the stochastic behavior and inherent volatility of traffic flow time series. Their parameters can be estimated directly from empirical data, making them data-driven yet mathematically expressive tools. While SDEs have been extensively applied in domains such as finance, their application in traffic flow modeling remains relatively limited, with notable use cases including short-term prediction \cite{tahmasbi2013modeling} and anomaly detection \cite{cheng2025traffic}. Volatility quantifies the statistical variation or uncertainity in a process over time statistically. It reflects the intensity of randomness or unpredictability inherent in the system. In the context of stochastic differential equations (SDEs), volatility is represented by the diffusion coefficient, which determines how strongly the Brownian motion (or noise) influences the process. In a highly volatile system a small fluctuation can lead to large fluctuations. In traffic systems, volatility can be characterized by the changing variance of traffic flow time series \cite{chen2012retrieval,zhang2014hybrid,kamarianakis2005modeling}.


Previous studies have demonstrated the potential of SDEs to capture complex traffic dynamics. For instance, models like Hull-White and Viscek have been utilized for short-term urban traffic volume forecasting. Despite their promise, SDEs remain underexplored in the transportation domain, and often lack integration with the underlying physics of traffic systems. For example, standard SDE formulations for traffic flow typically assume normally distributed outcomes, whereas empirical evidence suggests that traffic flow distributions are heavy-tailed, more accurately represented by Gamma, lognormal, or double-lognormal distributions. Although SDEs are not first-principles models, they can approximate real-world dynamics effectively when augmented with appropriate terms, features, and constraints. Stochasticity in traffic can also be understood in terms of heteroskedasticity, i.e., variability in variance over time—which has been modeled using GARCH-type approaches in hybrid stochastic models for urban traffic systems \cite{huang2017real, tsekeris2010short, sutarto2015parameter}. These developments highlight the potential of combining SDEs with heteroskedastic frameworks to achieve a more comprehensive understanding of urban traffic volatility.

\section{Data Description} \label{sec:empirical}
\begin{figure}[!htbp]\center
\includegraphics[width=1.0\linewidth]{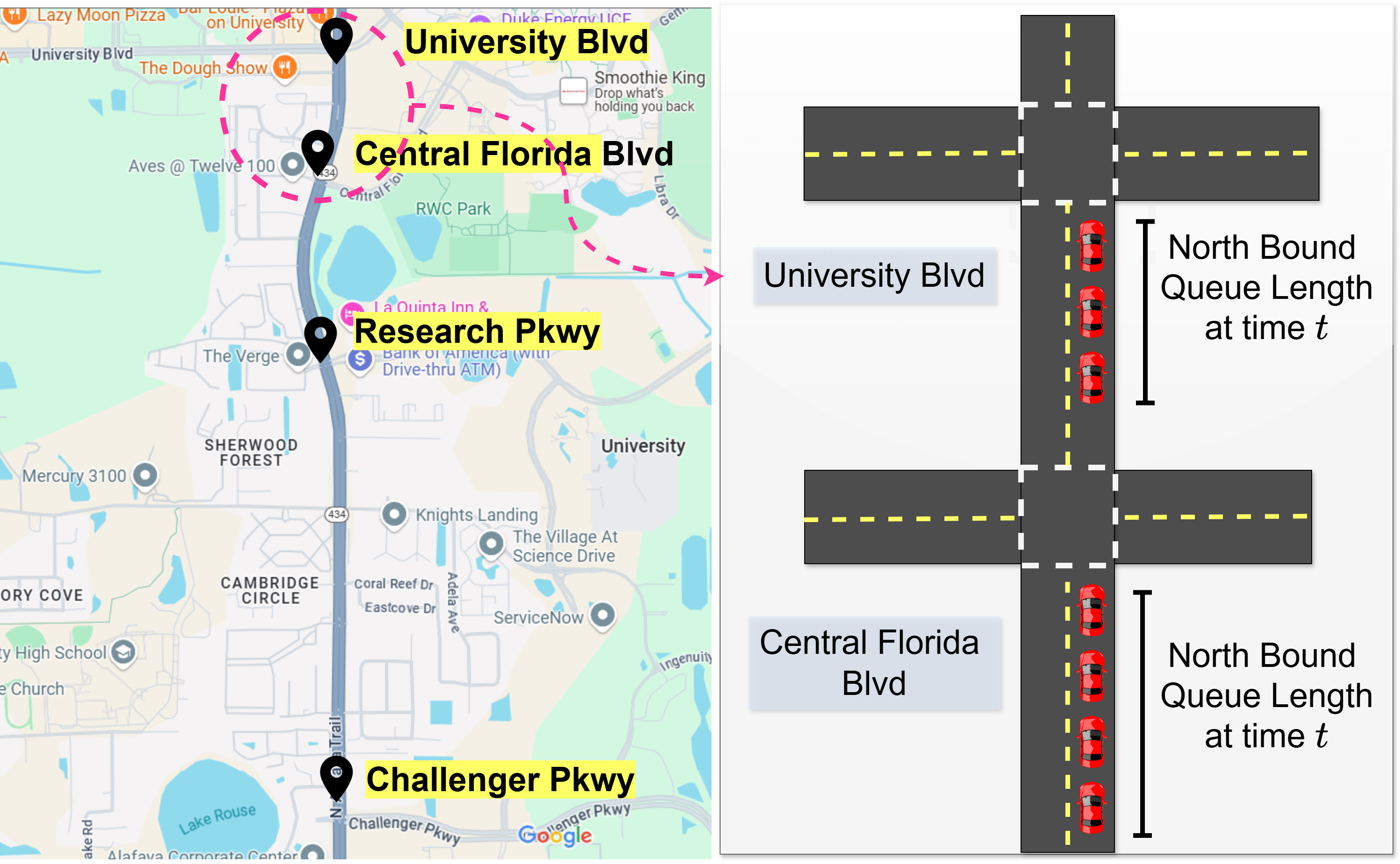}
\caption{Location of intersections on Alafaya corridor (left) and queue lengths are measured as a function of time. Each data point in the time series represents the maximum queue length recorded during a single signal cycle (right).}
\vspace{-4mm}
\label{fig:corridor}
\end{figure}

In this article, we analyze four signalized intersections along the Alafaya Trail (SR-434) corridor in East Orlando, FL (see the left panel of Figure~\ref{fig:corridor}). The selected intersections are Challenger Pkwy, Central Florida Blvd, Research Pkwy, and University Blvd. Queue length and traffic flow data were collected between December \(18,\; 2017\), and February \(14,\; 2018\). Queue lengths were measured using video cameras, while traffic flow data were obtained from inductive loop detectors. The InSync adaptive control system manages signal timing along this corridor \cite{insync_catalog}, which uses AI-based strategies to optimize queue lengths.

Our focus is on the cycle-to-cycle evolution of queue length dynamics. To that end, we construct a queue length time series by recording the maximum number of vehicles queued within each signal cycle (see the right panel of Figure~\ref{fig:corridor}). Because the control system is adaptive, cycle lengths vary between \(120\) and \(185\) seconds. To standardize the data, we aggregate queue lengths over uniform \(120\)-second intervals, which approximately correspond to the average cycle length. This time series provides a consistent surrogate for analyzing cycle-wise queue length variations.

Longer queue lengths are typically observed along the major roadway of a corridor. Consequently, adaptive signal control strategies primarily target queue management on these approaches \cite{shafik2017field}. Therefore, our analysis focuses on the northbound through (NT) movement, which is a major road along the Alafaya Trail corridor.

Traffic sensors are prone to inaccuracies caused by factors such as detector malfunctions, data storage errors, and adverse weather conditions. To address these issues, we applied a data cleaning procedure to identify and correct abnormal values. Along the corridor, sensors are installed between \(285\) and \(484\) feet upstream of the stop line to capture the full extent of vehicle queues. Given this setup, the maximum realistically detected queue length is \(40\) vehicles. Therefore, any queue lengths exceeding \(40\) vehicles are deemed unrealistic based on sensor placement. These values are removed and replaced with local averages computed using a five-point moving window. This window corresponds to a \(10\) minute interval. Additional details on data collection and quality control are available in \cite{rahman2021real}.

\section{Results and Discussions} \label{sec:results_discussion}
In this section, we will explore the scaling properties of queue length and traffic flow dynamics of the selected signalized intersections along the Alafya corridor (see the left panel of Figure~\ref{fig:corridor}. Based on the finding we  

\subsection{Power Law Scaling Characterization}
Here, we investigate the PSD in traffic flow and queue length time series. and find empirical evidence of a power-law relationship which indicate the self-similar nature of both traffic flow and queue length at a signalized intersection. 

\begin{figure}
    \centering
    \includegraphics[width=0.95\linewidth]{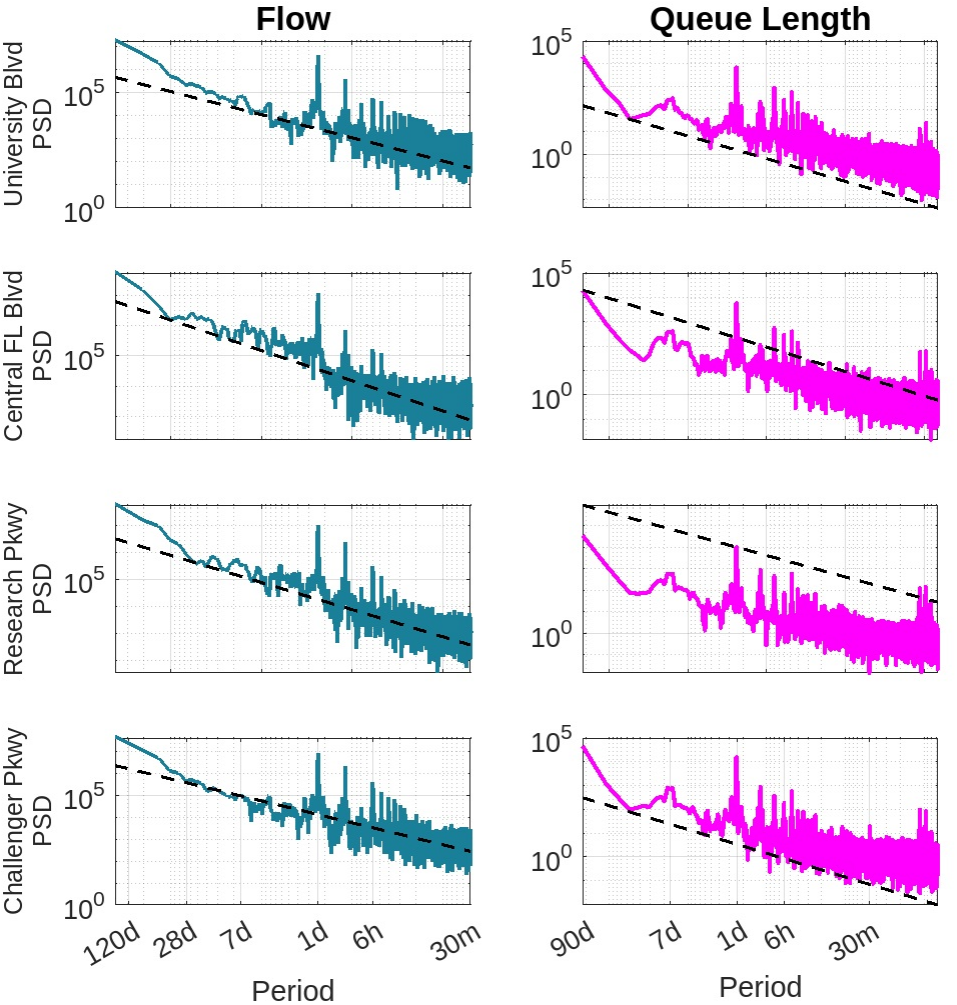}
    \caption{PSD of traffic flow and queue length of intersections illustrating \(1/f\) behavior found in the both time series. The dashed black line shows the \(-1\) slope reference}
    \label{fig:PSD_Flow}
\end{figure}

\noindent \textbf{PSD Structure:} Figure~\ref{fig:PSD_Flow} illustrates the PSD of the traffic flow time series and queue length time series of four signalized intersections considered here along the Alafaya corridor. The results show that the PSD of intersections exhibits \(1/f\)-type spectra. 

The traffic flow is sampled at a \(15\)-minute resolution. In contrast, queue length time series recorded at the same intersections have a \(2\) minute resolution, which is roughly equal to the traffic signal cycle duration. Traffic flow data was collected for \(120\) days, while queue length data were collected for \(90\) days. 

The x-axis of Figure~\ref{fig:PSD_Flow} represents the period (i.e., the reciprocal of frequency), ranging from 120 days for traffic flow and 90 days for queue length down to the period associated with respective Nyquiest frequency. providing a comprehensive view of temporal dynamics. The frequencies have been transformed into their corresponding periods to offer a clearer understanding of their physical relevance. Additionally, the spectrum was averaged over logarithmically spaced frequency bins to mitigate the influence of noise and reveal meaningful patterns. By analyzing the PSD of the queue length using the longest available dataset, we aim to ensure the robustness and reliability of the spectral characteristics.

Notably, both the traffic flow and queue length PSDs exhibit a \(1/f\)-like decay, despite being aggregated at different temporal granularities. 

This consistent power-law behavior suggests a scale-invariant structure in the underlying traffic dynamics across both macroscopic (flow) and microscopic (queue) subsystems.

\begin{figure}
        \includegraphics[width=\linewidth]{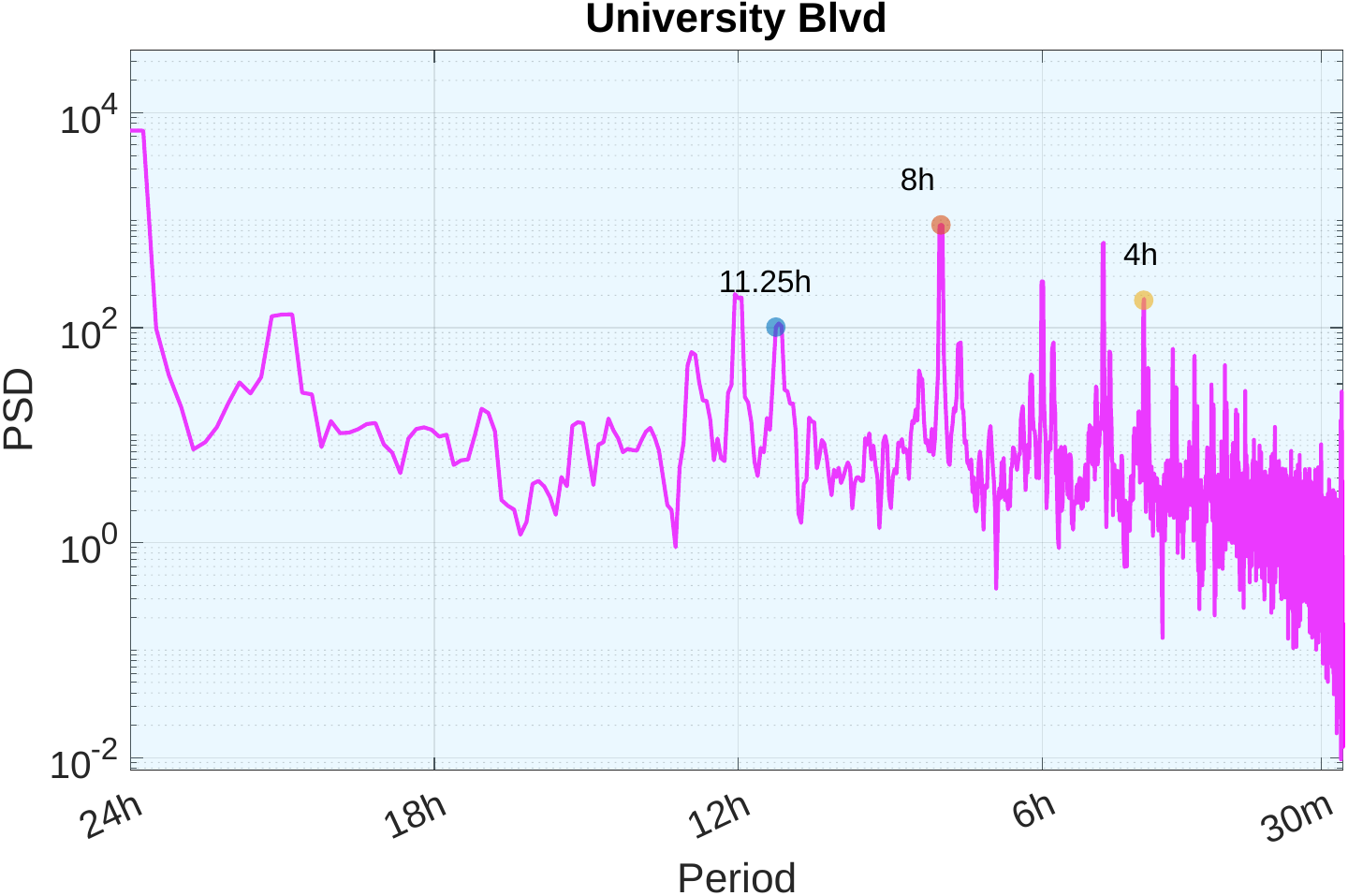} 
        \caption{PSD in semilog plot illustrates frequencies}
        \label{fig:zoomedPSD}
    \end{figure}

\noindent\textbf{Periodic Component: }We examined different regions of the spectrum by focusing on the University Blvd intersection. Figure~\ref{fig:zoomedPSD} shows a decreasing trend within the frequency range corresponding to $14$ days to $30$ minutes. Notably, sharp frequencies, representing periods with the highest energies, align with the system frequency for signalized intersections, including $24$ hr, $12$ hr, $8$ hr, $6$ hr, and $4$ hr, within this region. These frequencies denote periodic components of the system \cite{10070591}.

\subsection{Modeling Queue Length Dynamics} \label{sec:theoretical}


The observed \(1/f\) type behavior motivates the development of a stochastic model for queue length dynamics, which is presented in this section. We begin by modeling microscopic traffic flow as a point process, establish a connection between interarrival times and instantaneous traffic flow, and provide justification of the structure of the proposed stochastic differential equation (SDE).

\subsubsection{Vehicle Arrival as a Point-process}
A point process is a stochastic process consisting of a time series of discrete events occurring in continuous time. Similar modeling approaches are also found in fields like finance, where point processes describe high-frequency trading activity \cite{gontis2004multiplicative}. In transportation literature, traffic flow is often modeled as a point process, where each event corresponds to the arrival of a vehicle at a particular location \cite{lim2016traffic, brill1971point}.

Traffic arrival at an intersection will be modeled here as a point process. The arrival signal \( I(t) \) is defined as a sequence of pulses:

\begin{equation}
   I(t) = \sum_k a_k \delta(t - t_k) 
\end{equation}

where \( \delta(t) \) is the Dirac delta function, \( t_k \) denotes the arrival time of the \(k\)th event, and \( a_k \) is the contribution of that event to the signal. In the traffic context, \( a_k \) reflects the number of vehicles arriving at time \( t_k \), where \( a_k = 1 \) for a single vehicle and \( a_k > 1 \) for a vehicle platoon. When the pulse amplitude is constant, i.e., \( a_k = \bar{a} \), the process is fully described by the set of event times \( \{t_k\} \), or equivalently, the interarrival intervals \(  \tau_k = t_{k+1} - t_k  \). Stochastic models of these interarrival intervals define the dynamics of the point process. This formulation applies across multiple domains involving the flow of identical units, including electrons, photons, and vehicles \cite{KAULAKYS20001781}.

\begin{figure}
    \centering
    \includegraphics[width=0.95\linewidth]{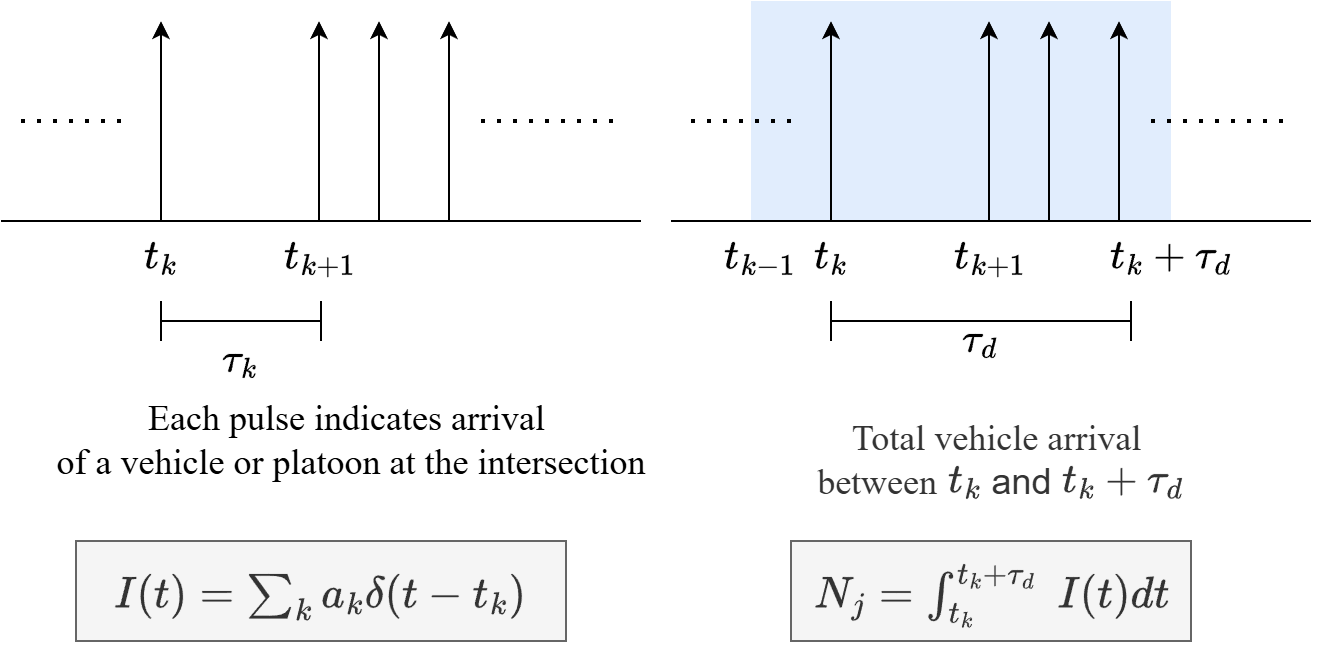}
    \caption{Schematic of modeling framework showing traffic arrival at an intersection a point process}
    \label{fig:enter-label}
\end{figure}

\subsubsection{Stochastic Interarrival Time and Instantenous Flow }

At free-flow conditions, the temporal headway between vehicles which also indicates interarrival time follows a Poisson distribution. This assumption arises from the randomness and independence of vehicle arrivals in low-density traffic regimes. However, in high-density or highly interactive traffic conditions, the temporal headway of individual vehicles no longer follows this simple distribution. Instead, it has been shown to follow a stochastic process with multiplicative noise resulting in a lognormal distribution \cite{wagner2004modelling}. Determining the overall distribution of interarrival times remains challenging, as it depends not only on stochastic interactions between vehicles but also on the underlying traffic demand, which is often periodic due to signal control and daily commuting patterns. Besides, signal coordination at a signalized corridor also determines vehicle arrival process. While simplified assumptions such as normally or Poisson-distributed arrivals are frequently used for analytical convenience, empirical studies suggest that the true nature of arrivals at signalized intersections is more complex \cite{LI2023128995}. Simulation-based investigations further reveal that interarrival times along corridors of signalized intersections are better represented by lognormal-type distributions \cite{boon2017arrival}.

Based on these properties, the interarrival time \( \tau_k \) between the \(k\)th and \((k+1)\)th vehicle is assumed to be controlled by a multiplicative noise:

\begin{equation}
    d\tau_k = \mu \tau_k\, dk + \sigma \tau_k\, dW_k
\end{equation}

where \( \mu \) and \( \sigma \) are the drift and diffusion coefficients, respectively, and \( W_k \) denotes a standard Wiener process indexed by vehicle number \(k\). This model is considered valid over short intervals where traffic demand and its variance can be assumed constant.

The instantaneous flow associated with the \(k\)th vehicle is defined as the reciprocal of the interarrival time, i.e., \( q_k = \tau_k^{-1} \). Applying Itô's lemma to this transformation, where \( q_k = f(\tau_k) = \tau_k^{-1} \), yields:

\begin{equation}
    dq_k = f'(\tau_k)\, d\tau_k + \frac{1}{2} f''(\tau_k)\, (d\tau_k)^2
\end{equation}

with \( f'(\tau_k) = -\tau_k^{-2} \), \( f''(\tau_k) = 2\tau_k^{-3} \), and \( (d\tau_k)^2 = \sigma^2 \tau_k^2\, dk \). Substituting these into the expression gives:

\begin{equation}
dq_k = -\frac{1}{\tau_k^2}(\mu \tau_k\, dk + \sigma \tau_k\, dW_k) + \frac{1}{\tau_k^3} \cdot \frac{1}{2} \cdot 2 \sigma^2 \tau_k^2\, dk    
\end{equation}

which simplifies to:

\begin{equation}\label{eq:instantFlow}
    dq_k = (\sigma^2 - \mu) q_k\, dk - \sigma q_k\, dW_k.    
\end{equation}

Thus, \( q_k \) is also influenced by multiplicative noise, but with modified drift coefficients. Also the sign of the diffusion coefficient is reversed. Equation~\ref{eq:instantFlow} indicates that if \(\tau_k\) is influenced by multiplicative noise, \(q_k\) is also influenced by multiplicative noise with different drift and diffusion coefficients.

Flow is measured within an observation window, making flow depend on the window. The instantaneous flow \(q_k\) is the inverse of the vehicular headway; therefore, it represents traffic flow in the shortest possible window. Now we can estimate the flow at any window using the instantaneous flow. Queue length can roughly be approximated as the accumulated traffic volume during the red cycle. So if we aggregate the instanteous at the cycle level, we get a rough maximum queue length at the cycle.



\subsubsection{Proposed SDE for Queue Length Dynamics}
This section bridges microscopic (instanteous flow) and macroscopic traffic flow (aggregated flow) dynamics. The influence of stochastic behavior at the microscopic level on macroscopic traffic patterns has been previously explored in the literature \cite{treiber2003memory}.

At signalized intersections, queue length emerges as a cumulative outcome of discrete vehicle arrivals aggregated over successive signal cycles. Assuming the queue is fully discharged during the green phase of each cycle, even if the queue length was not clear in the previous cycle, this will be captured by the stochastic term in the model. The main takeaway from the instantaneous model is that the noise is multiplicative, and in the cumulative process, the noise will still remain multiplicative. The total accumulated volume during the \(n\)th interval with each interval with duration \(T\) is given by:

\begin{equation}
X_n = \sum_{k:\, t_k \in [(n-1)T,\ nT]} q_k, \label{eq:cycle_flow}
\end{equation}


To describe maximum queue length over a cycle, we define a dynamic process \( Y_t \) that captures the cycle-wise evolution of maximum queue length. When interval length \(T\) corresponds to cycle length, i.e., approximately \(2\) min. we consider \(X_{n}\) becomes equivalent to \(Y_{t}\). Although arrival during the cycle is not exactly equal to maximum queue length becuase some vehicles can pass the intersection without any delay if arrive at green and some vehicle might at the intersection from the previous cycle. The approximation is still be can justified from the fact that the model considers stochasticity. Therefore, the contribution in the queue formation by left over vehicles from the previous cycle and the additional vehicle counts which will not be part of the queue due to arrival during green will be modeled by the stochastic component.    

This accumulation introduces temporal correlations due to the memory embedded in the integration process. To reflect this, we model \( Y_t \) as a fractional Brownian motion:

\begin{equation}
dY_t = \mu (\phi_t - Y_t)\, dt + \gamma_t Y_t\, dW_t^{H}, \label{eq:final}
\end{equation}

where \( W_t^H \) is a fractional Brownian motion with Hurst exponent \( H \in (0,1) \). The parameter \( H \) governs memory effects: \( H > 0.5 \) implies long-range dependence, \( H = 0.5 \) corresponds to memoryless Brownian motion, and \( H < 0.5 \) suggests anti-persistent behavior. Here, \( \phi(t) \) denotes the time-varying mean of the process, capturing the periodic nature of traffic demand. It reflects deterministic fluctuations in queue length caused by daily traffic patterns such as morning and evening peaks. The queue dynamics are thus decomposed into a deterministic component, governed by \( \phi(t) \), and a stochastic component representing random fluctuations. The procedure for estimating \( \phi(t) \) from empirical data will be detailed in Section~\ref{sec:mathsim}.  

Traffic flow over signal cycles also exhibits time-dependent volatility due to fluctuations in demand and signal coordination. To incorporate this, we model the volatility \( \gamma(t) \) using a mean-reverting Ornstein–Uhlenbeck-type square-root diffusion process:

\begin{equation} 
d\gamma(t) = \kappa \left( \bar{\gamma} - \gamma(t) \right) dt + \sigma_\gamma \, dW_t, \quad \gamma(t) \geq 0, \label{eq:volatility_process}
\end{equation}

where \( \bar{\gamma} \) is the long-run mean volatility, \( \kappa \) is the rate of mean reversion, and \( \sigma_\gamma \) is the volatility of the volatility process. The constraint \( \gamma(t) \geq 0 \) ensures physical realism by preventing negative volatility.\\

The general version of Equation~\ref{eq:final} can be written as 

\begin{equation}
    Y_{t} = F(Y_{t},\phi_{t})dt +G(Y_{t},\gamma_{t})dW_{t}^{H}
\end{equation}

Here, \(F(Y_{t},\phi_{t})\) represents the deterministic part of the dynamics, which is a mean reversion process. \(\phi_{t} = \sum_{i=1}^{n} A_{i}e^{j\omega_{i}t} \). Queue lengths of signalized corridors are either controlled by actuation control or adaptive control algorithms which adjusts green time by assigning a longer green time accroding to rush hours (actuated control) or assigns instanteous needs which introduces a negative feedback control or mean-reverting tendencies queue length dynamics which is captured in the model. 

The following two subsections will discuss simulation results. While Section~\ref{sec:mathsim} discusses the simulation process and values of the parameters, Section~\ref{sec:simres} describes the simulation results and compares with orginal time series and serves as valitation to the proposed model. 

\begin{figure}
    \centering
    \includegraphics[width=0.95\linewidth]{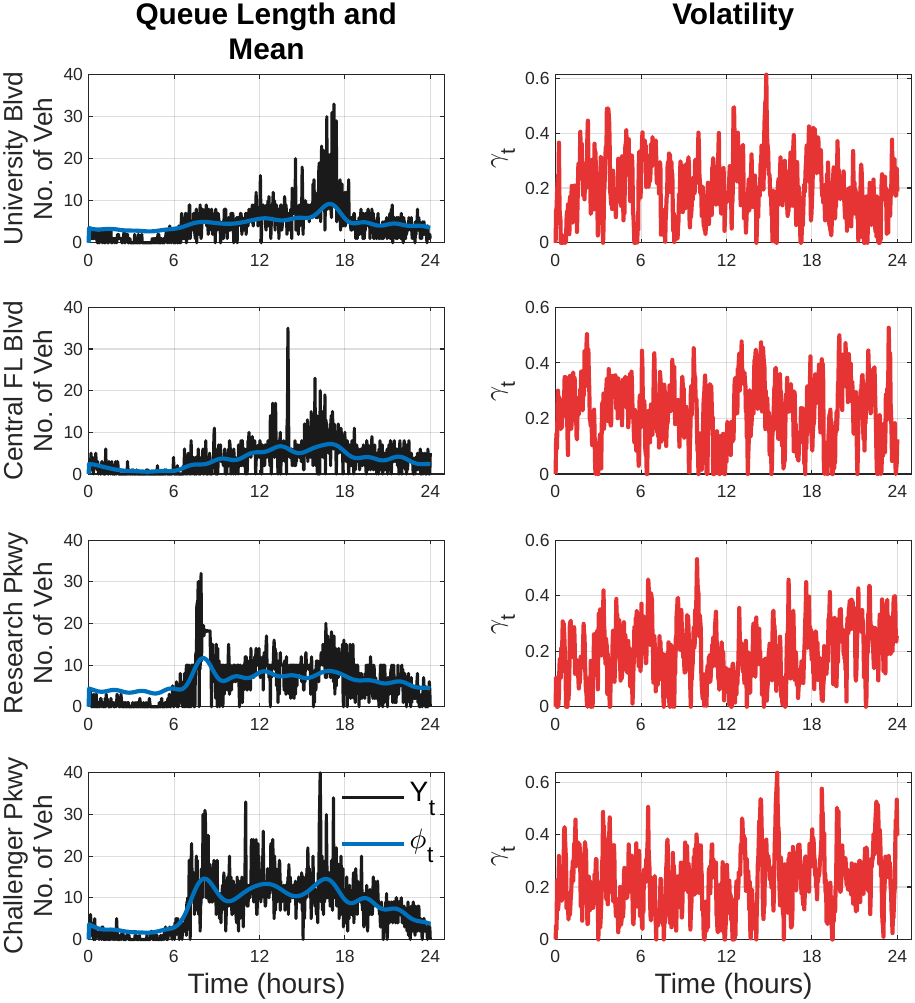}
    \caption{Illustration of time varying mean estimated from observation (left panel). The estimated volatility from the Ornstein-Uhlenbeck model is shown in the right panel.}
    \label{fig:mean}
\end{figure}

\subsection{Model Calibration and Validation} \label{sec:mathsim}

To replicate the observed queue dynamics at signalized intersections, we calibrated the parameters of a stochastic model through numerical optimization. The calibration process was executed individually for each intersection's queue time series using the following methodology.


\noindent \textbf{Seasonality Extraction:} To account for recurring temporal patterns, a seasonal trend $\phi_t$ was extracted via spectral decomposition. The empirical autocorrelation function (ACF) was used to estimate the dominant periodicity in the signal. A Fourier transform was subsequently applied to the zero-mean time series, and the top 12 harmonics surrounding the dominant frequency were retained to reconstruct the seasonal component. This reconstructed signal was smoothed and normalized to serve as a time-varying mean in the simulation process.

\noindent \textbf{Initial Guess:} The initial parameter vector for optimization are showed in Table~\ref{tab:initial_params}

\begin{table}[htbp]
\centering
\caption{Initial values for parameter calibration}
\begin{tabular}{llc}
\toprule
\textbf{Parameter} & \textbf{Description} & \textbf{Initial Value} \\
\midrule
$\mu$              & Mean-reversion speed of $X_t$           & 0.3 \\
$\gamma_0$         & Initial volatility level                & 0.1 \\
$\mu_\text{amp}$   & Amplitude of seasonal trend             & $\hat{\sigma}$ (empirical std. dev.) \\
$\mu_\text{base}$  & Baseline seasonal level                 & $\bar{x}$ (empirical mean) \\
$\kappa$           & Mean-reversion rate of $\gamma_t$       & 0.1 \\
$\sigma_\gamma$    & Volatility of volatility process        & 0.05 \\
$\bar{\gamma}$     & Long-term mean of volatility            & 0.2 \\
\bottomrule
\end{tabular}
\label{tab:initial_params}
\end{table}

where $\hat{\sigma}$ and $\bar{x}$ denote the empirical standard deviation and mean of the queue length time series, respectively.

\paragraph{Objective Function} The model parameters were estimated by minimizing the root-mean-square error (RMSE) between the simulated queue length trajectory and the empirical data. The optimization problem is formulated as
\[
\min_{\boldsymbol{\theta}} \; \text{RMSE} \left( \hat{x}_t(\boldsymbol{\theta}), x_t \right),
\]
where $\hat{x}_t(\boldsymbol{\theta})$ represents the simulated queue length generated by the stochastic differential model parameterized by $\boldsymbol{\theta}$, and $x_t$ denotes the observed empirical queue length.

\noindent \textbf{Optimization Algorithm: }The optimization was conducted using the Nelder–Mead simplex algorithm, as implemented in MATLAB's \texttt{fminsearch} routine. This algorithm is a derivative-free, local search method well-suited for problems with noisy or non-differentiable objective functions. The convergence tolerances were set to \texttt{TolFun} $=10^{-4}$ and \texttt{TolX} $=10^{-4}$.

\noindent \textbf{Simulation and Validation: }After obtaining the optimal parameter set, the model was simulated using the same initial condition as the empirical time series. The resulting synthetic trajectory was evaluated through visual inspection, empirical distribution comparison, and PSD analysis. These steps were used to validate both the statistical fidelity and dynamical consistency of the simulated queue lengths with respect to the real-world observations. Figure~\ref{fig:mean} illustrates the time-varying mean, \(\phi_{t}\) volatility \(\gamma_{t}\) estimated from the queue length time series.

\begin{figure}
    \centering
    \includegraphics[width=0.95\linewidth]{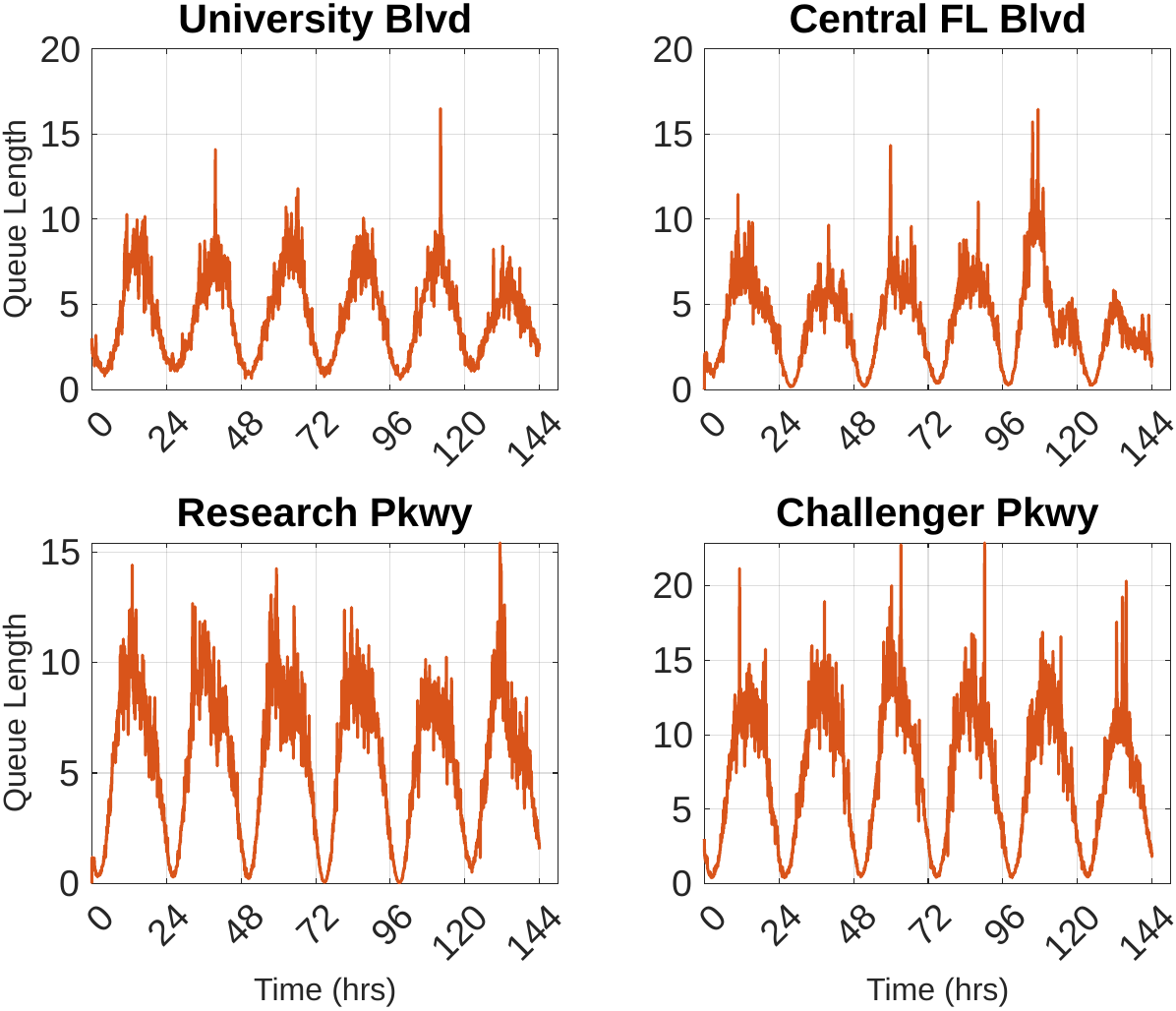}
    \caption{Simulated queue lengths for a week illustrates the proposed model is capable of capturing fluctuating behavior of queue length time series.}
    \label{fig:Sim_Queue}
\end{figure}

\begin{figure}
    \centering
    \includegraphics[width=0.95\linewidth]{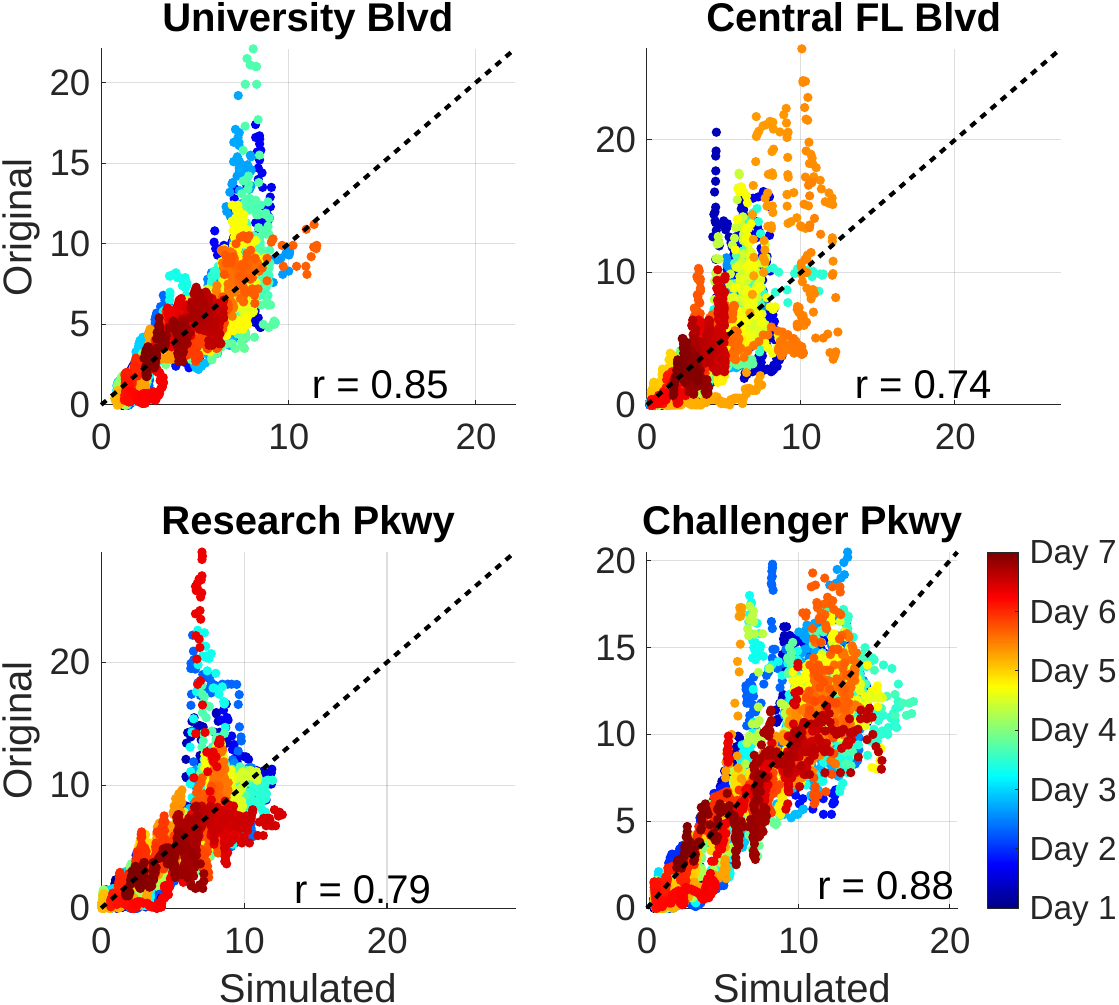}
    \caption{Scatter plot showing correlation between the simulated and original queue lengths. Here, simulation window was a week. the color bar signifies time evolution showing temporal information about deviation in the model.}
    \label{fig:Sim_scatter}
\end{figure}

\subsection{Simulation Results} \label{sec:simres}
\noindent \textbf{Simulated Queue Length: } This section presents the results from the simulation model and compares them with the observed data. Figure~\ref{fig:Sim_Queue} shows the simulated queue length time series over a one-week period using the proposed model. The results indicate that the model successfully captures the dynamics of queue length. To further validate the model, Figure~\ref{fig:Sim_scatter} illustrates the correlation between the simulated and observed time series. Given the stochastic nature of the model, an exact match is not expected. Nevertheless, the scatter plots demonstrate a strong agreement, with Pearson correlation coefficients \(r\) exceeding \(0.74\) in all cases. The correlation plots also visualize the temporal evolution of both series, helping identify whether outliers are associated with specific days or time windows. A reference line with a \(45^{\circ}\) slope is included; closer alignment of points along this line indicates better model performance. The color distribution appears balanced on both sides of the line, suggesting consistent model performance throughout the simulation period. Both the observed and simulated queue lengths were smoothed using a moving average with a window size of 10 in the figure.

\begin{figure}
    \centering
    \includegraphics[width=0.9\linewidth]{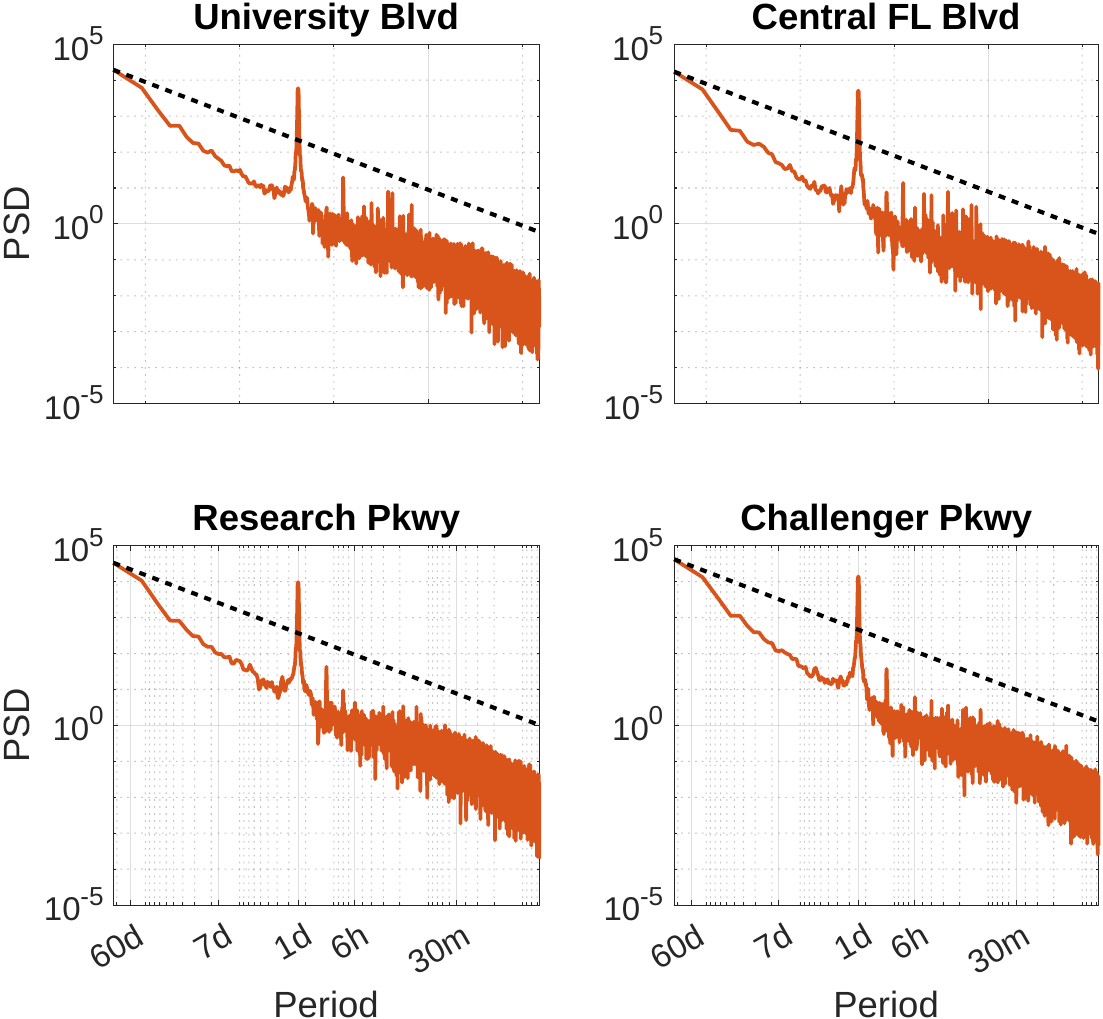}
    \caption{PSD of simulated queue lengths. The simulated PSD for all intersections displays a \(1/f\) type behavior. The dashed black line shows the \(-1\) slope reference}    \label{fig:Sim_PSD_Queue}
\end{figure}

\noindent \textbf{PSD Analysis: }Figure~\ref{fig:PSD_Flow} illustrates the PSD of the original queue length time series over a 60-day observation window, while Figure~\ref{fig:Sim_PSD_Queue} shows the PSD of the simulated time series. Both the original and simulated queue lengths exhibit a \(1/f\) type structure. The black dashed lines in the figures serve as references, indicating a slope of \(-1\). The simulation window was \(60\) days to observe the long term behavior of PSD.



\begin{figure}
    \centering
    \includegraphics[width=0.95\linewidth]{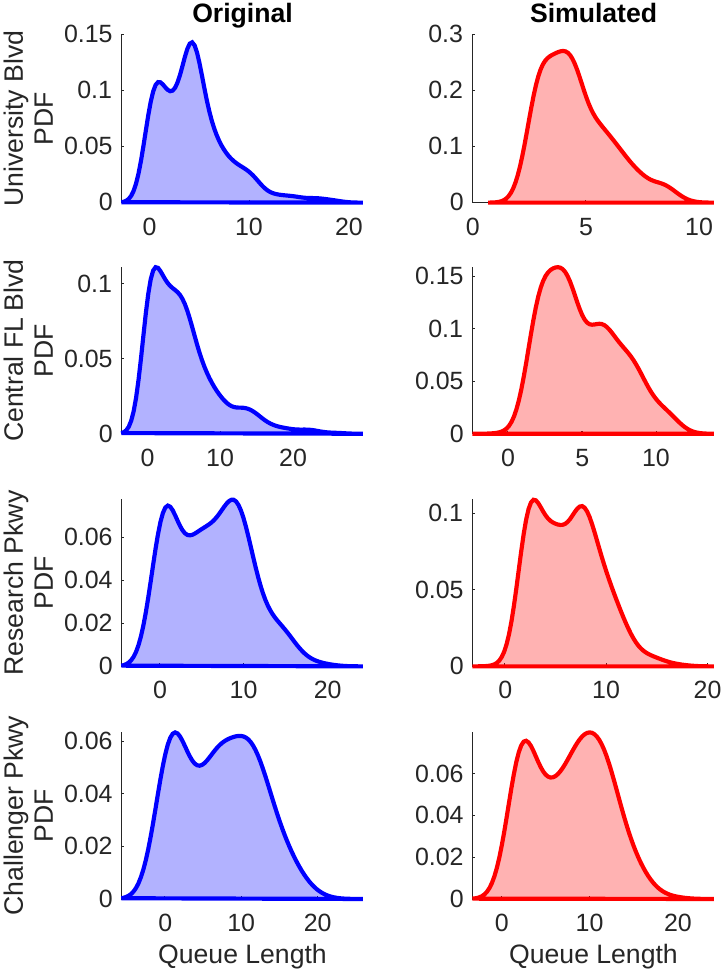}
    \caption{Comparison of original and simulated PDF for a single day.}
    \label{fig:Org_PDF_Queue}
\end{figure}

\noindent \textbf{PDF Comparison: } 
Figure~\ref{fig:Org_PDF_Queue} displays the PDF of the original time series and compares it with the PDF obtained from the simulation. The simulation window is one day. In both cases, the PDFs are similar. 


\section{Conclusion} \label{sec:conc}
This paper has proposed an SDE model to explain the dynamics of queue lengths at signalized intersections. The proposed model is novel in its ability to reproduce both the empirical PDF and the \(1/f\)-type behavior observed in the PSD of queue lengths. A key feature of the model is its incorporation of stochastic volatility, modeled using an Ornstein–Uhlenbeck (O-U) process. Although this choice has shown empirical effectiveness, alternative volatility models remain a promising avenue for future research. Understanding how the average behavior of \(\phi(t)\), which represents time-varying traffic demand, evolves with volatility could also yield more insights into the adaptivity of traffic dynamics. From an optimization standpoint, the Nelder–Mead simplex algorithm was employed due to its simplicity and ease of implementation; however, as a local optimizer, it may be limited in scope. Future studies could explore the use of more robust global optimization techniques. Another promising direction involves integrating the proposed SDE into physics-informed deep learning frameworks. While deep learning models often offer superior forecasting accuracy, equation-based models like the one presented here provide valuable interpretability and insight into the physical mechanisms governing traffic flow. These models not only capture stochastic variability and temporal patterns but also complement data-driven approaches by enhancing reliability and generalizability in hybrid learning systems.

\bibliographystyle{IEEEtran}
\bibliography{IEEE_fractal_ref}

\begin{thebibliography}{10}
\providecommand{\url}[1]{#1}
\csname url@samestyle\endcsname
\providecommand{\newblock}{\relax}
\providecommand{\bibinfo}[2]{#2}
\providecommand{\BIBentrySTDinterwordspacing}{\spaceskip=0pt\relax}
\providecommand{\BIBentryALTinterwordstretchfactor}{4}
\providecommand{\BIBentryALTinterwordspacing}{\spaceskip=\fontdimen2\font plus
\BIBentryALTinterwordstretchfactor\fontdimen3\font minus \fontdimen4\font\relax}
\providecommand{\BIBforeignlanguage}[2]{{%
\expandafter\ifx\csname l@#1\endcsname\relax
\typeout{** WARNING: IEEEtran.bst: No hyphenation pattern has been}%
\typeout{** loaded for the language `#1'. Using the pattern for}%
\typeout{** the default language instead.}%
\else
\language=\csname l@#1\endcsname
\fi
#2}}
\providecommand{\BIBdecl}{\relax}
\BIBdecl

\bibitem{li2025traffic}
Q.~Li, Z.~Liu, L.~Yao, H.~Li, G.~Xiong, and Y.~Zhang, ``Traffic signal coordinated control model for long arterial based on traffic flow spatiotemporal characteristics,'' \emph{Transportation Letters}, pp. 1--16, 2025.

\bibitem{maripini2023traffic}
H.~Maripini, A.~Khadhir, and L.~Vanajakshi, ``Traffic state estimation near signalized intersections,'' \emph{Journal of Transportation Engineering, Part A: Systems}, vol. 149, no.~5, p. 03123002, 2023.

\bibitem{zheng2019diagnosing}
G.~Zheng, X.~Zang, N.~Xu, H.~Wei, Z.~Yu, V.~Gayah, K.~Xu, and Z.~Li, ``Diagnosing reinforcement learning for traffic signal control,'' \emph{arXiv preprint arXiv:1905.04716}, 2019.

\bibitem{MEDINASALGADO2022100739}
\BIBentryALTinterwordspacing
B.~Medina-Salgado, E.~Sánchez-DelaCruz, P.~Pozos-Parra, and J.~E. Sierra, ``Urban traffic flow prediction techniques: A review,'' \emph{Sustainable Computing: Informatics and Systems}, vol.~35, p. 100739, 2022. [Online]. Available: \url{https://www.sciencedirect.com/science/article/pii/S2210537922000725}
\BIBentrySTDinterwordspacing

\bibitem{liu2009real}
H.~X. Liu, X.~Wu, W.~Ma, and H.~Hu, ``Real-time queue length estimation for congested signalized intersections,'' \emph{Transportation research part C: emerging technologies}, vol.~17, no.~4, pp. 412--427, 2009.

\bibitem{zhao2015ctm}
S.~Zhao, S.~Liang, H.~Liu, and M.~Ma, ``Ctm based real-time queue length estimation at signalized intersection,'' \emph{Mathematical Problems in Engineering}, vol. 2015, no.~1, p. 328712, 2015.

\bibitem{yin2018kalman}
J.~Yin, J.~Sun, and K.~Tang, ``A kalman filter-based queue length estimation method with low-penetration mobile sensor data at signalized intersections,'' \emph{Transportation Research Record}, vol. 2672, no.~45, pp. 253--264, 2018.

\bibitem{abewickrema2023multivariate}
W.~Abewickrema, M.~Yildirimoglu, and J.~Kim, ``Multivariate time-varying kalman filter approach for cycle-based maximum queue length estimation,'' \emph{Transportation research part C: emerging technologies}, vol. 154, p. 104238, 2023.

\bibitem{gu2024cycle}
Y.~Gu, S.~Razavi, and S.~Habibi, ``Cycle maximum queue length estimation: An integrated deep learning and adaptive neuro-fuzzy inference system framework (october 2024),'' \emph{IEEE Access}, 2024.

\bibitem{rahman2021real}
R.~Rahman and S.~Hasan, ``Real-time signal queue length prediction using long short-term memory neural network,'' \emph{Neural Computing and Applications}, vol.~33, pp. 3311--3324, 2021.

\bibitem{shirakami2023qtnet}
R.~Shirakami, T.~Kitahara, K.~Takeuchi, and H.~Kashima, ``Qtnet: Theory-based queue length prediction for urban traffic,'' in \emph{Proceedings of the 29th ACM SIGKDD Conference on Knowledge Discovery and Data Mining}, 2023, pp. 4832--4841.

\bibitem{sengupta2021tqam}
R.~Sengupta, Y.~Karnati, A.~Rangarajan, and S.~Ranka, ``Tqam: Temporal attention for cycle-wise queue length estimation using high-resolution loop detector data,'' in \emph{2021 IEEE International Intelligent Transportation Systems Conference (ITSC)}.\hskip 1em plus 0.5em minus 0.4em\relax IEEE, 2021, pp. 3313--3320.

\bibitem{wright2019neural}
M.~A. Wright, S.~F. Ehlers, and R.~Horowitz, ``Neural-attention-based deep learning architectures for modeling traffic dynamics on lane graphs,'' in \emph{2019 IEEE Intelligent Transportation Systems Conference (ITSC)}.\hskip 1em plus 0.5em minus 0.4em\relax IEEE, 2019, pp. 3898--3905.

\bibitem{ehlers2019traffic}
S.~F. Ehlers, ``Traffic queue length and pressure estimation for road networks with geometric deep learning algorithms,'' \emph{arXiv preprint arXiv:1905.03889}, 2019.

\bibitem{comert2021grey}
G.~Comert, Z.~Khan, M.~Rahman, and M.~Chowdhury, ``Grey models for short-term queue length predictions for adaptive traffic signal control,'' \emph{Expert Systems with Applications}, vol. 185, p. 115618, 2021.

\bibitem{vlahogianni2011temporal}
E.~Vlahogianni and M.~Karlaftis, ``Temporal aggregation in traffic data: implications for statistical characteristics and model choice,'' \emph{Transportation Letters}, vol.~3, no.~1, pp. 37--49, 2011.

\bibitem{feng2018better}
S.~Feng, X.~Wang, H.~Sun, Y.~Zhang, and L.~Li, ``A better understanding of long-range temporal dependence of traffic flow time series,'' \emph{Physica A: Statistical Mechanics and its Applications}, vol. 492, pp. 639--650, 2018.

\bibitem{yuan2017long}
P.~Yuan and X.~Lin, ``How long will the traffic flow time series keep efficacious to forecast the future?'' \emph{Physica A: Statistical Mechanics and its Applications}, vol. 467, pp. 419--431, 2017.

\bibitem{campari2002self}
E.~Campari and G.~Levi, ``Self-similarity in highway traffic,'' \emph{The European Physical Journal B-Condensed Matter and Complex Systems}, vol.~25, pp. 245--251, 2002.

\bibitem{musha19761}
T.~Musha and H.~Higuchi, ``The 1/f fluctuation of a traffic current on an expressway,'' \emph{Japanese Journal of Applied Physics}, vol.~15, no.~7, p. 1271, 1976.

\bibitem{takayasu19931}
M.~Takayasu and H.~Takayasu, ``1/f noise in a traffic model,'' \emph{fractals}, vol.~1, no.~04, pp. 860--866, 1993.

\bibitem{choi1995traffic}
M.~Choi and H.-Y. Lee, ``Traffic flow and 1/f fluctuations,'' \emph{Physical Review E}, vol.~52, no.~6, p. 5979, 1995.

\bibitem{paczuski1996self}
M.~Paczuski and K.~Nagel, ``Self-organized criticality and 1/f noise in traffic,'' in \emph{Traffic and granular flow}.\hskip 1em plus 0.5em minus 0.4em\relax World Scientific Singapore, 1996, p.~73.

\bibitem{nassab20051}
K.~Nassab, M.~Schreckenberg, S.~Ouaskit, and A.~Boulmakoul, ``1/f noise in a cellular automaton model for traffic flow with open boundaries and additional connection sites,'' \emph{Physica A: Statistical Mechanics and its Applications}, vol. 354, pp. 597--605, 2005.

\bibitem{wu2008long}
J.~Wu, H.~Sun, and Z.~Gao, ``Long-range correlations of density fluctuations in the kerner-klenov-wolf cellular automata three-phase traffic flow model,'' \emph{Physical Review E}, vol.~78, no.~3, p. 036103, 2008.

\bibitem{xue2015long}
Y.~Xue, L.-S. Jia, W.-Z. Teng, and W.-Z. Lu, ``Long-range correlations in vehicular traffic flow studied in the framework of kerner’s three-phase theory based on rescaled range analysis,'' \emph{Communications in Nonlinear Science and Numerical Simulation}, vol.~22, no. 1-3, pp. 285--296, 2015.

\bibitem{shang2006application}
P.~Shang, Y.~Lu, and S.~Kama, ``The application of h{\"o}lder exponent to traffic congestion warning,'' \emph{Physica A: Statistical Mechanics and its Applications}, vol. 370, no.~2, pp. 769--776, 2006.

\bibitem{shang2007fractal}
P.~Shang, M.~Wan, and S.~Kama, ``Fractal nature of highway traffic data,'' \emph{Computers \& Mathematics with Applications}, vol.~54, no.~1, pp. 107--116, 2007.

\bibitem{xu2010modeling}
N.~Xu, P.~Shang, and S.~Kamae, ``Modeling traffic flow correlation using dfa and dcca,'' \emph{Nonlinear Dynamics}, vol.~61, pp. 207--216, 2010.

\bibitem{li2007multifractal}
X.~Li and P.~Shang, ``Multifractal classification of road traffic flows,'' \emph{Chaos, Solitons \& Fractals}, vol.~31, no.~5, pp. 1089--1094, 2007.

\bibitem{shang2008detecting}
P.~Shang, Y.~Lu, and S.~Kamae, ``Detecting long-range correlations of traffic time series with multifractal detrended fluctuation analysis,'' \emph{Chaos, Solitons \& Fractals}, vol.~36, no.~1, pp. 82--90, 2008.

\bibitem{yue2010time}
J.~Yue, P.~Shang, and K.~Dong, ``Time-dependent hurst exponent in traffic time series,'' in \emph{2010 IEEE International Conference on Information Theory and Information Security}.\hskip 1em plus 0.5em minus 0.4em\relax IEEE, 2010, pp. 744--746.

\bibitem{wang2014multiscale}
J.~Wang, P.~Shang, and X.~Cui, ``Multiscale multifractal analysis of traffic signals to uncover richer structures,'' \emph{Physical Review E}, vol.~89, no.~3, p. 032916, 2014.

\bibitem{peng2010long}
S.~Peng, W.~Jun-Feng, T.~Tie-Qiao, and Z.~Shu-Long, ``Long-range correlation analysis of urban traffic data,'' \emph{Chinese Physics B}, vol.~19, no.~8, p. 080205, 2010.

\bibitem{krause2017importance}
S.~M. Krause, L.~Habel, T.~Guhr, and M.~Schreckenberg, ``The importance of antipersistence for traffic jams,'' \emph{Europhysics Letters}, vol. 118, no.~3, p. 38005, 2017.

\bibitem{he2016fractal}
H.-d. He, J.-l. Wang, H.-r. Wei, C.~Ye, and Y.~Ding, ``Fractal behavior of traffic volume on urban expressway through adaptive fractal analysis,'' \emph{Physica A: Statistical Mechanics and Its Applications}, vol. 443, pp. 518--525, 2016.

\bibitem{meng2009self}
Q.~Meng and H.~L. Khoo, ``Self-similar characteristics of vehicle arrival pattern on highways,'' \emph{Journal of Transportation Engineering}, vol. 135, no.~11, pp. 864--872, 2009.

\bibitem{perati2012self}
M.~R. Perati, K.~Raghavendra, H.~R. Koppula, M.~R. Doodipala, and R.~Dasari, ``Self-similar behavior of highway road traffic and performance analysis at toll plazas,'' \emph{Journal of transportation engineering}, vol. 138, no.~10, pp. 1233--1238, 2012.

\bibitem{laval2023self}
J.~A. Laval, ``Self-organized criticality of traffic flow: Implications for congestion management technologies,'' \emph{Transportation Research Part C: Emerging Technologies}, vol. 149, p. 104056, 2023.

\bibitem{tahmasbi2013modeling}
R.~Tahmasbi and S.~M. Hashemi, ``Modeling and forecasting the urban volume using stochastic differential equations,'' \emph{IEEE Transactions on Intelligent Transportation Systems}, vol.~15, no.~1, pp. 250--259, 2013.

\bibitem{cheng2025traffic}
Q.~Cheng, G.~Dai, B.~Ru, Z.~Liu, W.~Ma, H.~Liu, and Z.~Gu, ``Traffic flow outlier detection for smart mobility using gaussian process regression assisted stochastic differential equations,'' \emph{Transportation Research Part E: Logistics and Transportation Review}, vol. 193, p. 103840, 2025.

\bibitem{chen2012retrieval}
C.~Chen, Y.~Wang, L.~Li, J.~Hu, and Z.~Zhang, ``The retrieval of intra-day trend and its influence on traffic prediction,'' \emph{Transportation research part C: emerging technologies}, vol.~22, pp. 103--118, 2012.

\bibitem{zhang2014hybrid}
Y.~Zhang, Y.~Zhang, and A.~Haghani, ``A hybrid short-term traffic flow forecasting method based on spectral analysis and statistical volatility model,'' \emph{Transportation Research Part C: Emerging Technologies}, vol.~43, pp. 65--78, 2014.

\bibitem{kamarianakis2005modeling}
Y.~Kamarianakis, A.~Kanas, and P.~Prastacos, ``Modeling traffic volatility dynamics in an urban network,'' \emph{Transportation Research Record}, vol. 1923, no.~1, pp. 18--27, 2005.

\bibitem{huang2017real}
W.~Huang, W.~Jia, J.~Guo, B.~M. Williams, G.~Shi, Y.~Wei, and J.~Cao, ``Real-time prediction of seasonal heteroscedasticity in vehicular traffic flow series,'' \emph{IEEE Transactions on Intelligent Transportation Systems}, vol.~19, no.~10, pp. 3170--3180, 2017.

\bibitem{tsekeris2010short}
T.~Tsekeris and A.~Stathopoulos, ``Short-term prediction of urban traffic variability: Stochastic volatility modeling approach,'' \emph{Journal of Transportation Engineering}, vol. 136, no.~7, pp. 606--613, 2010.

\bibitem{sutarto2015parameter}
H.~Y. Sutarto, R.~K. Boel, and E.~Joelianto, ``Parameter estimation for stochastic hybrid model applied to urban traffic flow estimation,'' \emph{IET Control Theory \& Applications}, vol.~9, no.~11, pp. 1683--1691, 2015.

\bibitem{insync_catalog}
{Rhythm Engineering}, ``Insync traffic signal optimization catalog,'' \url{https://rhythmtraffic.com/insync/#DownloadCatalog}, accessed: 2025-06-14.

\bibitem{shafik2017field}
M.~S.~I. Shafik, ``Field evaluation of insync adaptive traffic signal control system in multiple environments using multiple approaches.''

\bibitem{10070591}
S.~Das, S.~Mustavee, S.~Agarwal, and S.~Hasan, ``Koopman-theoretic modeling of quasiperiodically driven systems: Example of signalized traffic corridor,'' \emph{IEEE Transactions on Systems, Man, and Cybernetics: Systems}, vol.~53, no.~7, pp. 4466--4476, 2023.

\bibitem{gontis2004multiplicative}
V.~Gontis and B.~Kaulakys, ``Multiplicative point process as a model of trading activity,'' \emph{Physica A: Statistical Mechanics and its Applications}, vol. 343, pp. 505--514, 2004.

\bibitem{lim2016traffic}
K.~W. Lim, W.~Wang, H.~Nguyen, Y.~Lee, C.~Cai, and F.~Chen, ``Traffic flow modelling with point processes,'' in \emph{Proceedings of the 23rd World Congress on Intelligent Transport Systems}, 2016, pp. 1--12.

\bibitem{brill1971point}
E.~A. Brill, ``Point processes arising in vehicular traffic flow,'' \emph{Journal of Applied Probability}, vol.~8, no.~4, pp. 809--814, 1971.

\bibitem{KAULAKYS20001781}
\BIBentryALTinterwordspacing
B.~Kaulakys and T.~Meškauskas, ``Models for generation 1/f noise,'' \emph{Microelectronics Reliability}, vol.~40, no.~11, pp. 1781--1785, 2000. [Online]. Available: \url{https://www.sciencedirect.com/science/article/pii/S0026271400000858}
\BIBentrySTDinterwordspacing

\bibitem{wagner2004modelling}
P.~Wagner, ``Modelling traffic flow fluctuations,'' \emph{arXiv preprint cond-mat/0411066}, 2004.

\bibitem{LI2023128995}
\BIBentryALTinterwordspacing
M.~Li, J.~Tang, Q.~Chen, and Y.~Liu, ``Traffic arrival pattern estimation at urban intersection using license plate recognition data,'' \emph{Physica A: Statistical Mechanics and its Applications}, vol. 625, p. 128995, 2023. [Online]. Available: \url{https://www.sciencedirect.com/science/article/pii/S0378437123005502}
\BIBentrySTDinterwordspacing

\bibitem{boon2017arrival}
M.~Boon, ``Arrival processes at traffic intersections,'' 2017.

\bibitem{treiber2003memory}
M.~Treiber and D.~Helbing, ``Memory effects in microscopic traffic models and wide scattering in flow-density data,'' \emph{Physical Review E}, vol.~68, no.~4, p. 046119, 2003.

\end{thebibliography}

\begin{IEEEbiography}[{\includegraphics[width=1in,height=1.25in,clip,keepaspectratio]{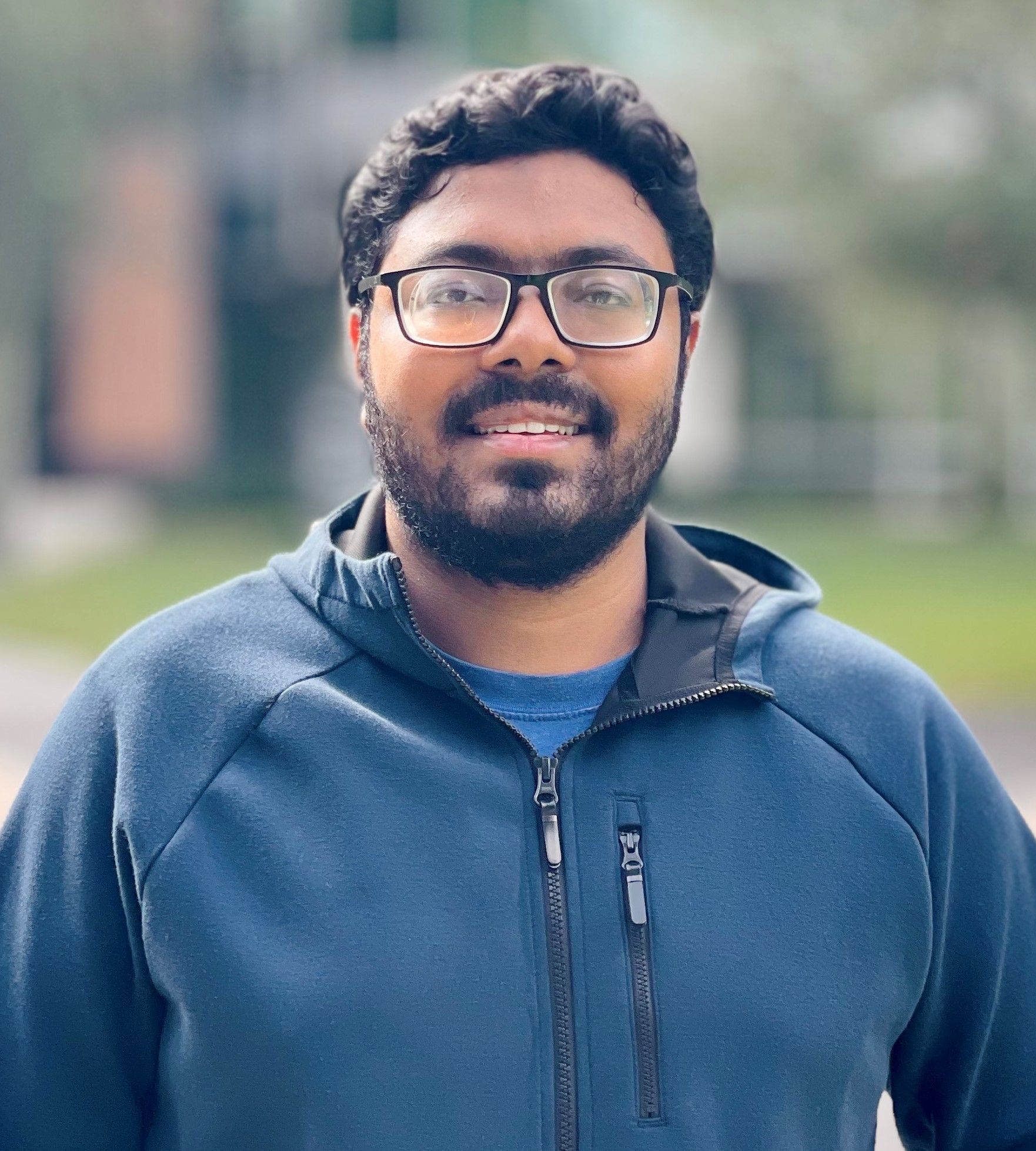}}]%
{Shakib Mustavee}is a Post-doctoral Researcher in the Civil, Environmental, and Construction Engineering (CECE) Department at the University of Central  Florida (UCF) from where he also completed his PhD in 2024. His research interests include data-driven modeling of real-world complex systems.
\end{IEEEbiography}
\vskip 0pt plus -1fil
\begin{IEEEbiography}[{\includegraphics[width=1in,height=1.25in,clip,keepaspectratio]{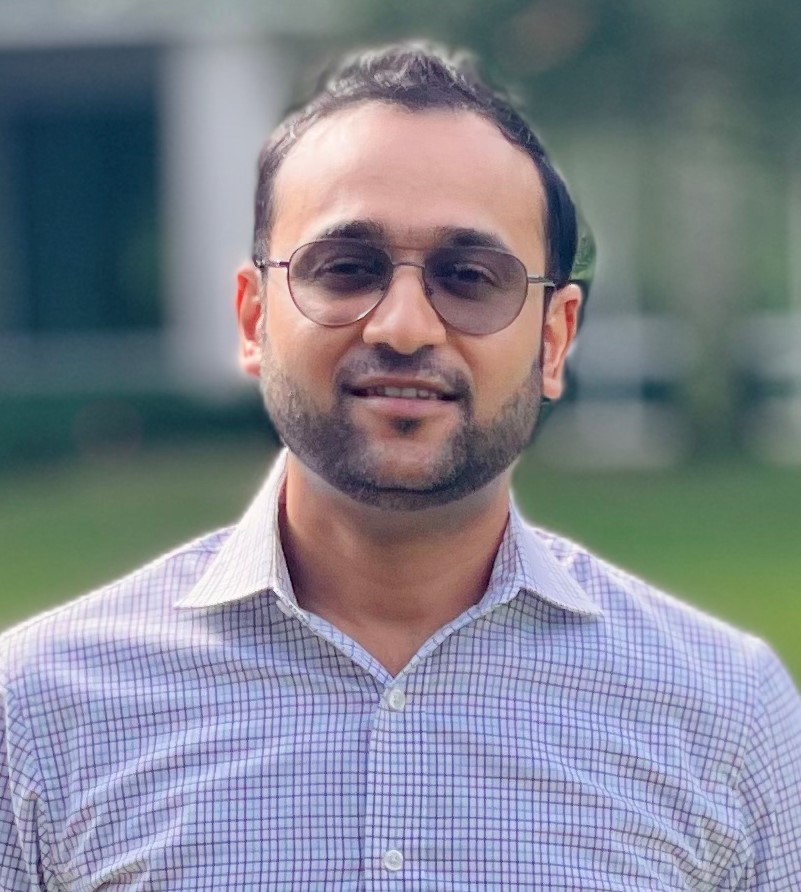}}]%
{Shaurya Agarwal}is an Associate Professor in the CECE Department at the University of Central Florida (UCF). He completed his post-doctoral research at New York University and his Ph.D. in Electrical Engineering from the University of Nevada, Las Vegas (UNLV). He is a senior member of IEEE and currently serves as an associate editor of IEEE Transactions on Intelligent Transportation Systems.
\end{IEEEbiography}

\begin{IEEEbiography}[{\includegraphics[width=1in,height=1.25in,clip,keepaspectratio]{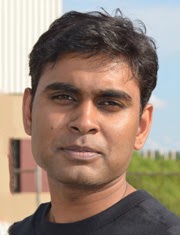}}]%
{Arvind Singh}is an Associate Professor in the CECE Department at the University of Central Florida (UCF). He completed his post-doctoral research and Ph.D. in Civil Engineering from the University of Minnesota, Minneapolis, MN. His research focuses on sediment transport, landscape evolution modeling, turbulence, multiscale analysis of hydro-geomorphological processes, nonlinear dynamics.
\end{IEEEbiography}


\end{document}